\documentclass
{amsart}
\usepackage{amssymb,amsmath,graphicx}
\theoremstyle{definition}

\theoremstyle{remark}

\numberwithin{equation}{section}

\begin{document}
\title [String Cosmologies] {Interfacing Friedman and
String Cosmologies}
\author{J. Towe}
\address{Department of Physics, The Antelope Valley College, Lancaster, CA 93536}%
\email{jtowe@avc.edu}\
\begin{abstract}
Compactification and symmetry breaking events in theories of
heterotic superstrings implicitly indicate five inflation events.
Scales of $10^{9}$ ly and larger are modeled in terms of these
events. The proposed model accounts qualitatively for recent
observations (of galactic clusters that lie on "thin,..
semi-spherical shells") and motivates a calculation that accounts
(approximately) for the observed number of galaxies.


\end{abstract} \maketitle
\section {Five Inflation Events}\label{S:intro}
Traditionally it was believed that the distribution of mass-energy
was homogeneous and isotropic on scales of $10^{8}$ light years and
larger [J. A. Wheeler, 1974]. In recent years however, the
homogeneity of the large-scale distribution has been challenged. 
After surveying more than 5000 galaxies, J. P. Huchra and M. Geller
concluded that "galactic clusters lie on thin, sharply defined
semi-spherical shells that surround vast, empty voids." They
compared these phenomena with the "bubble-like structures that
constitute the foam in a kitchen sink" [J.P. Huchra, 1986]. Thus,
while the distribution of mass-energy is isotropic (established by
observations indicating that the background radiation is isotropic),
it is not homogeneous. These observational results have challenged
traditional theories, but it is argued here that they can be
understood in terms of a supersymmetric gauge theory that indicates
multiple symmetry breaking events: i.e. multiple inflation events.
\par
In the theory of superstrings, one avoids anomalies and
contradictions if the dimension of spacetime is 10. Moreover, one
rids the ground state of tachyons and introduces spacetime
supersymmetry through GSO projections. Periodic GSO projections
involve Ramon boundary conditions, which produce a Type IIA theory,
containing spin-1 fields and fermions. Anti-periodic GSO projections
involve Nevu-Schwarz boundary conditions, which produce a Type IIB
theory, containing scaler fields and fermions. Neither theory is
physical. The first is not chiral and the second contains no gauge
fields. One can solve this problem however, by combining the two
theories into a mixed or 'heterotic' theory. One approach to this
involves compactification of 16 dimensions from the 26 dimensional
bosonic theory, which imposes a symmetry $E_{8}XE_{8}$ onto the
10-spacetime that corresponds to the heterotic theory.
\par
One $E_{8}$ symmetry characterizes the observable sector while the
other characterizes the hidden sector (where coupling with gauge
fermions in the observable sector is strictly in terms of
supergravity). When 10-spacetime is compactified to produce
$M_{4}XK_{6}$, each $E_{8}$ is broken to yield SU(5)XSU(3). Guth
inflation [A. Guth, 1981] is a theoretical device that maintains
large scale flatness by compensating (by rapid expansion) for the
fattening of the vacuum that is due to a Higgs event (an event in
which Higgs scalers are swallowed by other particles, imparting mass
to the latter and therefore breaking the initial symmetry). Clearly
however, a Guth inflation event is also called for by a
compactification that increases mass-energy density; i.e. fattens
the vacuum by reducing the number of spacetime dimensions. Thus one
Guth event is generated by compactification itself. A second Guth
event is generated by the breaking of $E_{8}$ to yield SU(5)XSU(3).
A third inflation event is generated by the breaking of the other
$E_{8}$ symmetry to yield a second version of SU(5)XSU(3). And
finally fourth and fifth inflation events are generated by the
breaking of the two SU(5) symmetries. In the proposed theory then, a
single Guth event is replaced by five such events.
\par
The proposed additional inflation events complement the original
shock wave that is attributed to the big bang with secondary,
tertiary etc. shock waves as prescribed by Huygens' principle.
Specifically, the first inflation event is regarded as spherically
symmetric. The second inflation event is associated with secondary
shock waves that emanate from the points of the initial, spherically
symmetric wave front. Similarly the third inflation event is
associated with tertiary shock wavelets, emanating from the points
of the secondary wave fronts etc. Because these are inflation
events, scales on the secondary, tertiary etc. wavelets are much
larger than the scales on the wave fronts from which they emanate.
In this context, it is argued that the five inflation events that
are produced by compactification and symmetry breaking account for
the large scale. 
A hypothesis relating numbers of galaxies to inflation events will
now be considered.

\section {A Theoretical Number of Galaxies}\label{S:intro}
The original inflation event, is attributed here to compactification
of 16 dimensions onto a torus and is interpreted as the big bang. It
is argued that this event formed a radially symmetric distribution
of mass-energy that can be identified as a pre-galaxy. It is assumed
that this initial pre-galaxy is a generic phenomenon that
subsequently became part of every galaxy. The second inflation
event; i.e. the first Huygens event is associated with the fattening
of the vacuum that corresponds to compactification from 10 spacetime
dimensions to $M_{4}XK_{6}$. It is argued that this inflation event
produced a cluster of pre-galaxies; that the third inflation event;
i.e. the second Huygens event (that associated with the second
compactification from 10 spacetime dimensions to $M_{4}XK_{6}$);
e.g. the compactification associated with the hidden sector
generated a supercluster of pre-galaxies; that the fourth inflation
event; i.e. the third Huygens event (that associated with the
breaking of one SU(5) symmetry) generated a cluster of
superclusters; and finally, that the fifth inflation event; i.e. the
fourth Huygens event (that associated with the breaking of the
second SU(5)) generated a supercluster of superclusters. In the
context of this hypothesis and of boundary conditions that are
determined from observations of local clusters, one can calculate
the number of galaxies that populates the universe as modeled.
\par
Based upon observations of local clusters, one adopts the following
boundary conditions: 1. A typical galaxy means that this galaxy is
contained in a typical basic cluster; i.e. a basic cluster
containing an average number of galaxies. To ascertain the nature of
a typical basic cluster in the local group of thirty five or so
galaxies, it is observed that the large and small Magellanic clouds
form a pair of satellites about the Milky Way, and that the
Andromeda Nebula is at the center of a basic cluster that includes
satellite galaxies M-32, NGC-205, M-33, NGC-47 and NGC-185. Based
upon this and several similar observations, it is concluded that the
average number of galaxies constituting a basic cluster in the local
group is five; 2. \textit{Separations of galaxies are about ten
times the diameters of the galaxies; separations of galactic
clusters are about ten times the diameters of the clusters;
separations of superclusters are about ten times the diameters of
the superclusters etc.}
\par
The iterative counting process, which is based upon the above stated
hypothesis and boundary conditions, is as follows: It is determined
(from observation) that the typical galaxy is about
c$t_{0}$=$10^{5}$ light years (ly) in diameter. The diameter
c$t_{N}$ of the global state that has resulted from the $N^{th}$
inflation event is, by provision number 2, designated:
\begin{equation}
ct_{N}=ct_{0}10^{N }: N=0,1,2,3,...,
\end{equation}
where the N=0 state corresponds to a single, typical galaxy.
\par
Because five galaxies populate the typical basic cluster and
because, up to a scale of about $10^{9}$ ly, the number of galaxies
can be enlisted as units in terms of which to express volume (up to
this scale, galaxies appear to fill a volume rather than to populate
the surfaces of semi-spherical shells), one can establish the
following equation to describe the N=1 cluster of galaxies:
\begin{equation}
(4/3)\pi R_{1}^{3}=5,
\end{equation} which implies that
\begin{equation}
R_{1}=1.06.
\end{equation}
From 2.1 the radius of this galactic cluster in light years is about
\begin{equation}
ct_{1}=ct_{0}10=10^{5}(10)ly=10^{6}ly.
\end{equation}
\par
One now determines the approximate radius (as a number of galaxies)
of the N=2 state. Given the counting device that is italicized
above, it is argued that the radius of the N=2 state (the separation
of the N=1 cluster and a second cluster is ten times the diameter of
the local cluster), in terms of a number of galaxies, is given by
\begin{equation}
10(diameter(local cluster))=10(2(1.06))\cong 21.
\end{equation}
The radius in light years of a typical N=2 state is (consulting
equation 2.1)
\begin{equation}
ct_{2}=ct_{0}10^{2}=10^{5}(10^{2})ly=10^{7}ly.
\end{equation}

Proceeding in this way, one determines the radius (as an approximate
number of galaxies) of the N=3 state. By analogy with the above
calculations, the radius of the N=3 state, in terms of a number of
galaxies, is given as a multiple of a typical separation of N=2
systems:
\begin{equation}
10(diameter(N=2 state))=10(2(21))=420
\end{equation}
The radius of the N=3 state in light years is (according to 2.1)
about
\begin{equation}
ct_{3}=ct_{0}10^{3}ly=10^{5}10^{3}ly=10^{8}ly.
\end{equation}
\par
One now determines the radius (as an approximate number of galaxies)
of the N=4 state. The radius of this state, in terms of a number of
galaxies, is given by
\begin{equation}
10(diameter(N=3 state))=10(2(420))=8400.
\end{equation}
The radius in light years of the N=4 state is, according to equation
2.1, given by
\begin{equation}
ct_{4}=ct_{0(10^{4})}=(10^{5})(10^{4})ly=10^{9}ly.
\end{equation}
\par
Finally, one determines the radius (as an approximate number of
galaxies) of the N=5 state. The radius of this state, in terms of a
number of galaxies, is given by

\begin{equation}
10(diameter(N=4 state))=10(2(8400))=168000.
\end{equation}
Thus, since galactic clusters populate the surfaces of semi-spheres
on scales larger then $10^{8}$ ly, the number of galaxies in the N=5
state is (summing areas of opposite semi-spherical shells) given by
\begin{equation}
4\pi (R_{4})^{2}=4(3.14)(168000)^{2}=3.54X10^{11}.
\end{equation}
The radius in light years of the N=5 state is, according to equation
2.1,
\begin{equation}
ct_{5}=ct_{0}(10^{5})ly=10^{10}ly.
\end{equation}
\par
In summary, the proposed large scale hypothesis is based upon
compactification and symmetry breaking events that are associated
with string theory. It is argued that five inflation events result
from these, and that the resulting inflations account qualitatively
for the radial inhomogeneity of the large scale structure as
recently observed. Based upon this structure moreover, and upon
locally determined boundary conditions, an iterative calculation
predicts the approximate number of galaxies that is indicated by
observation. 

\end{document}